\documentclass[11pt,a4paper]{article}
\usepackage{longtable}
\usepackage{amsmath}
\usepackage{amssymb}
\usepackage{graphicx}
\usepackage{bm}
\usepackage{footmisc}
\usepackage{afterpage}
\usepackage{float}
\usepackage{lscape}
\usepackage{longtable}
\usepackage{float}
\usepackage{slashed}
\usepackage{morefloats}
\usepackage{afterpage}
\usepackage{mathrsfs}  

\newcommand*{\La}{\mathcal{L}}
\newcommand*{\no}{\noindent}
\newcommand*{\bea}{\begin{eqnarray}}
\newcommand*{\eea}{\end{eqnarray}}
\newcommand*{\be}{\begin{equation}}
\newcommand*{\ee}{\end{equation}}
\newcommand*{\pd}{\partial}
\newcommand*{\pdm}{\pd_{\mu}}
\newcommand*{\pdn}{\pd_{\nu}}

\newcommand*{\pref}[1]{(\ref{#1})}

\newcommand*{\mn}{{\mu\nu}}

\newcommand*{\nn}{\nonumber}

\newcommand{\bma}{\begin{pmatrix}}
\newcommand{\ema}{\end{pmatrix}}

\newcommand*{\op}{{{\cal O}}}
\newcommand*{\la}{\left\langle}
\newcommand*{\ra}{\right\rangle}
\newcommand*{\ms}{\phantom{\mu}}
\newcommand\refeq[1]{\stackrel{(\ref{#1})}{=}}

\newcommand*{\igc}{(g^c)} 
\newcommand*{\rcl}{R^c} 
\newcommand*{\rclB}{(R^c)} 
\newcommand*{\nabcl}{\nabla^c} 
\newcommand*{\inabcl}{(\nabla^c)} 
\newcommand*{\dAcl}{{\Box^c}} 
\newcommand{\irtcl}{(R^c)} 
\newcommand{\gdcl}{\sigma^c} 
\newcommand{\gdclB}{(\sigma^c)} 
\newcommand{\evA}{X} 
\newcommand{\evB}{Y} 
\newcommand{\chrisCl}{(\Gamma^c)} 
\newcommand{\prop}{\pi^c} 

\usepackage[square,comma,sort&compress,numbers]{natbib}

\title{Exploratory applications of the Fr\"ohlich-Morchio-Strocchi mechanism in quantum gravity}

\author{Axel Maas$^1$, Markus Markl$^{1,2}$, Michael M\"uller$^{1,3,4}$\\[0.5cm]
$^1$ Institute of Physics, NAWI Graz, University of Graz,\\
Universit\"atsplatz 5, A-8010 Graz, Austria\\
$^2$ Fusion@\"OAW, Institut f\"ur Theoretische Physik - \\
Computational Physics, Technische Universit\"at Graz, \\ Petersgasse 16, 8010
Graz, Austria \\
$^3$ Department of Physics, University of Guelph, \\
Guelph, Ontario, Canada, N1G 2W1 \\
$^4$ Perimeter Institute for Theoretical Physics, \\
Waterloo, Ontario, Canada, N2L 2Y5
}

\begin{document}

\maketitle

\begin{abstract}
A manifestly diffeomorphism-invariant approach to canonical quantum gravity requires to use composite operators. These can be considered to be bound states of matter and/or gravitons, intrinsically non-perturbative objects. An analytical approach to determine the properties of such bound states could be the Fröhlich-Morchio-Strocchi mechanism. We explore the necessary technology by applying it to various $n$-point functions, including geon propagators and black-hole-particle vertices.
\end{abstract}

\maketitle

\section{Introduction}

Strict independence of human choices requires that observables are manifestly and non-perturbatively invariant under choices of coordinate systems or gauges \cite{Berghofer:2021ufy,Frohlich:1980gj,Frohlich:1981yi,Lavelle:1995ty,Maas:2017wzi}. In quantum gravity this translates to invariance of observables under diffeomorphisms. In loop quantum gravity this is pushed to the extreme by using exclusively diffeomorphism-invariant variables \cite{Ashtekar:2004eh,Ashtekar:2011ni}. An alternative approach is to still use diff\-eo\-mor\-phism-dependent quantities, like the metric or vierbein, as dynamical variables, but to determine eventually diffeomorphism-invariant observables, just as is done in ordinary quantum gauge theories. However, this approach is substantially hampered by the complexity such observables entail, as these are necessarily described by composite operators.

However, observationally quantum gravity is dominated at long distances by classical physics, at least for the values of Newton's constant and the cosmological constant relevant to our universe. In such a situation the Fr\"ohlich-Morchio-Strocchi (FMS) mechanism \cite{Frohlich:1980gj,Frohlich:1981yi,Maas:2017wzi} from non-gravitational theories is a possibility to unravel composite operators using analytic methods. The FMS mechanism can indeed be transferred, at least in principle, to canonical quantum gravity \cite{Maas:2019eux}, if the theory can be formulated in terms of a gauge theory quantizable by a path integral. The latter is supported by results from scenarios like asymptotic safety \cite{Niedermaier:2006wt,Reuter:2012id,Reuter:2019byg,Bonanno:2020bil} as well as dynamical triangulation approaches \cite{Ambjorn:2012jv,Loll:2019rdj}.

Thus, the present purpose will be to make the conceptual ideas of \cite{Maas:2019eux} more concrete, though still in an exploratory manner. These explorations are a necessary next step before being able to go to real quantitative calculations. As it turns out, arguably expected, that quantum gravity harbors a couple of additional challenges in applying the FMS mechanism in comparison to ordinary flat-space quantum field theory. To this end, we investigate here how the FMS approach could be applied to a selection of $n$-point functions with $n\le4$ in sections \ref{s:cosmo}-\ref{s:behss}.

A particularly important intermediate step is the question of a suitable gauge to apply the FMS mechanism. This will be considered in section \ref{s:gauge}. In a sense, this defines a possible framework for the quantitative application of the FMS mechanism in quantum gravity. It also discusses in more detail how to usefully define the fluctuation field in this context.

We summarize our experiences in section \ref{s:sum}. We find in particular that it is, already from a classical point of view, far from obvious how to formulate certain questions in a manifestly diffeomorphism invariant way. However, we find that at the quantum level these need to be phrased very carefully to obtain meaningful answers.

\section{Setup and gauge fixing}\label{s:gauge}

We consider here a path-integral approach, in which we include Einstein-Hilbert gravity, as well as a single scalar particle. The quantization is performed using a path-integral approach, with the action
\bea
S&=&\int d^4x\sqrt{\det(-g)}\left(\frac{1}{2\kappa}\left(R+l\right)+g^\mn \pd_\mu\phi\pd_\nu\phi+m\phi^2\right)\\
&=&\int d^4x\sqrt{\det(-g)}\left(-\frac{1}{8\kappa}
g^{\alpha \beta } g^{\mu \nu } \left(-4 \partial_{\nu }\partial_{\beta }g_{\alpha \mu } + 4 \partial_{\nu }\partial_{\mu }g_{\alpha \beta }\right.\right.\nn\\
&&\left.\left.+ g^{\rho \sigma } \left(2 \partial_{\nu }g_{\beta \sigma } \partial_{\rho }g_{\alpha \mu } - 3 \partial_{\rho }g_{\alpha \mu } \partial_{\sigma }g_{\beta \nu }\right.\right.\right.\nn\\
&&\left.\left.\left.+ \partial_{\mu }g_{\alpha \beta } (\partial_{\nu }g_{\rho \sigma } - 4 \partial_{\sigma }g_{\nu \rho }) + 4 \partial_{\beta }g_{\alpha \mu } \partial_{\sigma }g_{\nu \rho } \right)\hspace{-3mm}\phantom{\frac{1}{1}}-4l\right) \right.\nn\\
&&\left.\ +g^\mn \pd_\mu\phi\pd_\nu\phi+m\phi^2\right)\nn
\eea
\no where $\kappa$ and $l$ are suitable combinations of Newton's constant and the cosmological constant, $R$ is the curvature scalar, $g$ the metric, $\phi$ a scalar field and $m$ the mass the scalar particle would have at tree-level in flat space-time. Note that for scalars the covariant derivatives become ordinary ones.

We assume for the moment that a path-integral quantization is possible, where ultraviolet stabilization occurs due to a dynamical effect like asymptotic safety \cite{Niedermaier:2006wt,Reuter:2012id,Reuter:2019byg,Bonanno:2020bil}, possibly supplied with further terms in the action  \cite{Schaden:2015wia,Asaduzzaman:2019mtx,Catterall:2009nz}. This assumption will affect any quantitative results in the following, but our qualitative steps will be unaltered. As integration variables, we will use the metric and the scalar field, though the vierbein or other quantities would be possible as well. Again, the following could be conducted analogously for any other such choice. It is here chosen entirely for technical simplicity, and may not be suitable eventually \cite{Hehl:1976kj,Maas:2019eux}. This requires to chose the integration range. Here, we will restrict to all compatible metrics, which satisfy $\det g<0$ and have signature $+2$, to allow for local transformations to flat space.

The basic idea \cite{Maas:2019eux} of the FMS mechanism \cite{Frohlich:1980gj,Frohlich:1981yi} is to formulate observables in a manifestly diffeomorphism-invariant way, and afterwards expand them in quantum fluctuations around a classical metric. This needs to be done after gauge-fixing, and is thus conceptually different than in any background-field approach. The question of how to perform the split is far from obvious at the moment. However, defining
\be
g_\mn=g_{\mu\rho}^c \tilde{\gamma}^\rho_\nu=g_{\mu\rho}^c\left(\delta^\rho_\nu+(g^{c})^{\rho\sigma}\gamma_{\sigma\nu}\right)=g_\mn^c+\gamma_\mn\label{fmssplit},
\ee
\no where $g_\mn^c$ is a fixed classical metric, shows that a linear and a product split are both equal. This implies that $\det\tilde\gamma>0$, and it cannot alter the signature of $g_\mn^c$. Thus, also $\gamma_\mn$ cannot change the signature. The linear shift allows to switch to $\gamma$ as integration variable in the path integral without a Jacobian. Physically, $\gamma$ measures the change of distances $ds^2$ compared to the reference distance $(ds^c)^2=g_\mn^c d x^\mu dx^\nu$. It is important to note that only $g_\mn$ can be used to lower and raise indices, while trying to do so with $g_\mn^c$ yields an error of order $\gamma$.

While these are so far exact statements, to apply the FMS mechanism we will moreover assume that $\gamma$ is small, or likewise that $\tilde\gamma$ is close to the unit matrix. The linear split is technically most convenient in the following, and we will use it exclusively from here on.

By construction, it was required that $(g^c)^{\mn}$ is the inverse of $g_\mn^c$. As a consequence, the same does not hold true for $\gamma_\mn$ and $\gamma^\mn$, and it follows
\be
\delta^\mu_\rho=g^\mn g_{\nu\rho}=((g^c)^\mn+\gamma^\mn)(g_{\nu\rho}^c+\gamma_{\nu\rho})=\delta^\mu_\rho+(g^c)^\mn\gamma_{\nu\rho}+g^c_{\nu\rho}\gamma^\mn+\gamma^\mn\gamma_{\nu\rho}\nn.
\ee
\no This implies the linear relation
\be
\gamma^\mn\left(g_{\nu\rho}^c+\gamma_{\nu\rho}\right)+(g^c)^\mn\gamma_{\nu\rho}=0\label{invgamma}
\ee
\no which is formally implicitly solved by a Dyson-like equation
\be
\gamma^\mn=-(g^c)^{\mu\sigma}\gamma_{\sigma\rho}g^{\rho\nu}\label{dyson}.
\ee
\no While in a numerical calculation an equation like \pref{invgamma} can be solved, this is not always possible in an analytical approach.

However, here we are mainly concerned with a perturbative treatment, which implies that $\gamma$ can be considered small in comparison to $g^c$. We can therefore expand \pref{dyson} in the fluctuation field, yielding
\be
\gamma^\mn=-(g^c)^{\rho\nu}(g^c)^{\mu\sigma}\gamma_{\sigma\rho}+(g^c)^{\rho\nu}(g^c)^{\mu\sigma}\gamma_{\sigma\alpha}(g^c)^{\alpha\beta}\gamma_{\beta\rho}-...\label{giexp}
\ee
\no and thus obtain the necessary desired inverse fluctuation field $\gamma^\mn$ in terms of the fluctuation field. Note, however, that this immediately implies the presence of an infinite number of tree-level terms in a perturbative expansion\footnote{As it has been found that in the FMS approach divergencies from ordinary perturbation theory can cancel at the level of observables \cite{Maas:2020kda}, it is an interesting question whether these additional terms may actually help to make a perturbative approach to quantum gravity predictive. This will have to await the technology to do loop calculations in this setup.}.

As noted, the split needs to be done after gauge-fixing. With the aim of technical simplification, it turns out to be useful that the full gauge condition $C[g]=0$ is also fulfilled by the classical metric, $C[g^c]=0$. This is certainly always possible. In addition, a simple form of the classical metric will also be helpful in the following calculations. Hence, the gauge condition chosen should allow for a simple explicit form. Given the usual expressions for flat and (anti-)de Sitter metric, a suitable choice is the Haywood gauge
\be
g^{\mu\rho}\pd_\rho g_{\nu\mu}=\pd^\mu g_{\nu\mu}=0\nn.
\ee
\no This implies for classical metrics obeying the Haywood gauge condition
\be
g^{\mu\rho}\pd_\rho\gamma_{\nu\mu}=0=(g^c)^{\mu\rho}\pd_\rho\gamma_{\nu\mu}+\gamma^{\mu\rho}\pd_\rho\gamma_{\nu\mu}\label{haywood}.
\ee
\no Inserting \pref{giexp} into \pref{haywood} therefore implies that at fixed order in $\gamma$ there is always an additional term of one order higher in $\gamma$ in the gauge condition than the order to which one works, and by which it is violated. That is, however, not different from perturbative treatments in ordinary gauge theories.

Inserting the Haywood gauge condition in the action and performing the usual Faddeev-Popov construction finally yields the gauge-fixed path integral
\bea
Z&=&\int{\cal D}(g\bar{c}c) e^{iS}\nn\\
S&=&\int d^4x\sqrt{\det(-g)}\times\nn\\
&&\times\left(\frac{1}{2\kappa}\left(R+l\right)+\bar{c}_\nu\pdm\left(D^{\nu\mu}_{\ms\ms\sigma}+ D^{\mu\nu}_{\ms\ms\sigma}\right) c^\sigma+g^\mn \pd_\mu\phi\pd_\nu\phi+m\phi^2\right)\nn\\
R&=&- \tfrac{1}{4} g^{\alpha \beta } g^{\mu \nu } \Bigl(-4 \partial_{\nu }\partial_{\beta }g_{\alpha \mu } + 4 \partial_{\nu }\partial_{\mu }g_{\alpha \beta } \Bigr.\nn\\
 && \Bigl. + g^{\rho \sigma } \bigl(\partial_{\mu }g_{\alpha \beta } \partial_{\nu }g_{\rho \sigma } + \partial_{\rho }g_{\alpha \mu } (2 \partial_{\nu }g_{\beta \sigma } - 3 \partial_{\sigma }g_{\beta \nu })\bigr)\Bigr)\nn\\
D_{\mu\nu}^{\ms\ms\rho}&=&\pdm\delta_\nu^{\ms \rho}-\Gamma^\rho_{\ms\mu\nu}\nn \label{eqn:cov_deriv}\\
\Gamma^\rho_{\ms\mu\nu}&=&\frac{1}{2}g^{\rho\sigma}\left(\pdm g_{\rho\nu}+\pdn g_{\mu\rho}-\pd_\rho g_\mn\right)\nn.
\eea
\no Note that at this point the expression remains exact, and $g$ is still the full quantum metric, though restricted to obey the Haywood gauge condition \pref{haywood}.

\section{One-point functions and cosmology}\label{s:cosmo}

While formally the split \pref{fmssplit} can be performed for any $g^c$, the aim is, of course, to simplify calculations. Thus, the split \pref{fmssplit} is a purely technical tool. Hence, a good choice is essential to make progress. This needs to be accompanied also by a suitable choice of gauge-fixing condition, as otherwise additional technical complications will arise.

This requires a few reflections upon the anticipated behavior of the theory. Quantum gravity is based on events, rather than coordinates, as elementary objects. Any path integral formulation with a fully diffeomorphism-invariant action and measure does not introduce any preference of events, and all events are equal. As a consequence, no quantity which depends on a single event can be anything but zero, except for scalars which can be non-zero, but event-independent. Especially any expectation value of the metric vanishes without gauge fixing. Space-time can only be characterized by (covariantly constant) densities of curvature scalars. These are primarily quantities like the curvature scalar, the Kretschmann scalar, the Weyl scalar, and so on.

Choosing $g^c$ such that it reproduces the observed values of these scalars implies the absence (or cancellation) of quantum corrections for them. It is therefore a well-motivated possibility. Of course, since these scalars are diffeomorphism-invariant, there is still considerable freedom in this choice. Imposing in addition the gauge condition \pref{haywood} singles out finally a particular form\footnote{Up to possible Gribov copies.}. Here, this will be flat space-time, as the simplest case, and the de Sitter metric in the Cartesian Roberston-Walker chart, which yields a diagonal metric. This leads to the observed cosmological constant, and the corresponding curvature scalar.

When considering our universe, it would appear that the Friedmann-Lem\^aitre-Robertson-Walker (FLRW) metric should also be a good choice given how well it works in cosmology. And of course, it is certainly possible. However, in this case quantities like the curvature scalar are not constant, implying the need for quantum corrections. This appears odd at first sight. But there is a reason behind this. In this case a special event, the big bang, is introduced, and the dependence of , e.\ g., the curvature scalar is given in terms of the eigentime since this special event. Hence, this choice does not respect that all events are equal. This equality would then needed to be restored by quantum corrections.

This leads immediately to the question how something like cosmology and the existence of a universe could be described in this setup. To answer this requires to pose the question how to describe a universe in terms of manifestly diffeomorphism-invariant quantities. This is a surprisingly non-trivial question.

In fact, the closest possibility is probably the following point of view, which considers a universe more like a scattering process\footnote{In fact, the scattering aspect only comes in because one wants to talk about time evolution. Taking into account that this is part of the problem, the situation is actually more akin to a many-body problem in ordinary quantum field theory. Cosmology relates to quantum gravity thus in a similar fashion as, e.\ g., a neutron star to QCD.}. For simplicity, assume that the universe contains only two particles\footnote{It is actually unknown to the authors if a lower limit for the particle content of a universe to drive cosmology exists.}, described by two diffeomorphism-invariant operators $O_i$. Given then the matrix element
\be
\la O_1^\dagger(X) O_2^\dagger (X) O_1(Y) O_2(Z)\ra\nn
\ee
\no the corresponding cross section can be used to describe the probability of the development of a universe. To this end, determine the cross-section as a function of a space-like geodesic distance $r=\la \sigma(Y,Z)\ra$ at fixed time-like geodesic distances $\tau=\la \sigma(X,Y)\ra=\la \sigma(X,Z)\ra$, with the geodesic distances determined as in \cite{Maas:2019eux}. This describes how likely this universe's size $r$ is as a function of the elapsed time-like duration $\tau$. If there would be more than two particles, the largest pairwise distance $r$ would provide the size of the universe. Note that in this way no event is preferred as the big bang. The properties of the universe in this way follow indeed  as an expectation value and characterize an average universe created from the employed matter Lagrangian. An actual universe, like ours, is then the consequence of a measurement process, which is governed by this probability. Thus, in this way it is calculable how likely a universe like ours actually is. Quantities like the curvature of the universe would then be obtained from distributions of geodesic distances, and are not an inherent observable derived directly from operators like the curvature scalar. This is, as it must be, since these quantities depend on the other matter.

Of course, a practical calculation for our universe, with its estimated $10^{80}$ particles, is not feasible (yet). However, expecting a quasi-classical behavior, as one sees also for cross sections of many particles in ordinary particle physics, could lead to the hope that the probability would peak around the observed behavior. In addition, many-body techniques could perhaps also be developed in this case.

Note that it is important in this context to carefully distinguish between the classical metric for the expansion point of the FMS mechanism and the effective behavior of the universe. The prior is determined solely by the Lagrangian and the parameters of the theory. The latter is also affected by the number of particles which are put into the universe, and thus its initial condition. The former may, e.\ g., be flat Minkowski space-time while the latter be what is expected in classical general relativity with an FLRW metric.

\section{Two-point functions and the geon}\label{s:geon}

When describing individual particles, this is usually done using their propagators. In light of the previous discussion, such propagators in the present setup describe the properties in the absence of other matter. Thus, just using the propagator will only be a suitable approximation in our universe if the propagation distance is small compared to distances to other particles\footnote{But this is also true in ordinary QFT, just that in this case no dynamical description of the universe is possible.}. Thus the quantities to be discussed will be expectation values like $\la\op(X)\op(Y)\ra$.

As described in section \ref{s:gauge}, this will be done using the FMS mechanism to lowest order. As a consequence, the argument will be just the classical geodesic distance $\gdcl(X,Y)$ between the events $X$ and $Y$. Hence, the propagator of the scalar particle
\be
\la \phi(X)\phi(Y)\ra=D_\phi(\gdcl(X,Y)))\label{scprop}
\ee
\no will be just the corresponding scalar propagator in the respective classical space-time, e.\ g.\ in flat space-time just the ordinary one. At higher orders these two features will change due to quantum fluctuations. Calculations in Euclidean Dynamical Triangulation indeed find that the full propagator, including its functional dependence, differs at most slightly from \pref{scprop} \cite{Dai:2021fqb}.

It is therefore interesting to consider the simplest objects which necessarily deviate from the flat space-time behavior at this order. To avoid the complications with spin, we consider only spinless objects. To obtain a non-trivial behavior requires also that the corresponding operator contains the metric. There are, of course, an infinite number of these. The simplest one is arguably the curvature scalar $R(X)$. Of course, all other scalar operators will mix with it. However, we will operate on the prejudice from quantum-field theory that the leading-order contribution of the simplest operator will have overlap with the ground state.

Thus, to lowest order in the FMS mechanism the physical, composite object has a propagator of the form discussed above,
\be
\la R(X)R(Y)\ra=D_R(\gdcl(X,Y))\nn.
\ee
\no To obtain this form explicitly, it is necessary to apply the split in \eqref{fmssplit} to the curvature scalars and to order the resulting terms in powers of $\gamma$. As the curvature scalar involves also the inverse metric, we use again \pref{invgamma}. This yields
\bea
\langle R(X)R(Y) \rangle &=& \bigg\langle R_X^{(0)}R_Y^{(0)} + \left[R_X^{(0)}R_Y^{(1)} + R_X^{(1)}R_Y^{(0)}\right] \nonumber\\
  &&+\left[R_X^{(1)}R_Y^{(1)} + R_X^{(0)}R_Y^{(2)} + R_X^{(2)}R_Y^{(0)} \right] + O\left(\gamma^3\right)\bigg\rangle \nn
\eea
\no up to second order, where superscripts denote the order in $\gamma$ and subscripts the event at which the object is evaluated. Now for both of our cases the classical curvature scalar is either vanishing or constant. At the same time, because of invariance under the choice of events, the propagator cannot depend on a single event. Hence, the only contribution which can be non-constant at this order is $\langle R_X^{(1)}R_Y^{(1)} \rangle$.

Thus, this requires the corresponding expression for the Ricci scalar to first order in the FMS expansion. The second order accurate approximation of the Ricci scalar can be expressed as follows:
\bea
R &=& g^{\mu \nu}R_{\mu \nu} = \left(\igc^{\mu \nu} + \gamma^{\mu \nu}\right)R_{\mu \nu} \nn\\
&=& \left(\igc^{\mu \nu} - \igc^{\mu \alpha}\igc^{\nu \beta}\gamma_{\alpha \beta} + \igc^{\mu \alpha}\igc^{\beta \rho}\igc^{\nu \sigma}\gamma_{\alpha \beta}\gamma_{\rho \sigma} + O\left(\gamma^3\right)\right) \nonumber\\
&& \times \left( \rcl_{\mu \nu} + R_{\mu \nu}^{(1)} + R_{\mu \nu}^{(2)} +O\left(\gamma^3\right)\right)\nn\\
R^{(0)} &=& \rcl = \igc^{\mu \nu}\rcl_{\mu \nu} \nn\\
R^{(1)} &=& \igc^{\mu \nu}R_{\mu \nu}^{(1)}- \igc^{\mu \alpha}\igc^{\nu \beta}\gamma_{\alpha \beta}\rcl_{\mu \nu} \nn\\
R^{(2)} &=& \igc^{\mu \nu}R_{\mu \nu}^{(2)}- \igc^{\mu \alpha}\igc^{\nu \beta}\gamma_{\alpha \beta}R_{\mu \nu}^{(1)}+\igc^{\mu \alpha}\igc^{\beta \rho}\igc^{\nu \sigma}\gamma_{\alpha \beta}\gamma_{\rho \sigma} \rcl_{\mu \nu}\nn
\eea 
The explicit expressions for the quantities $R^{(1)}$ and $R^{(2)}$, in terms of $\gamma_{\mu\nu}$, are determined in appendix \ref{a:geon}. For the rest of section \ref{s:geon} we will assume $\igc^{\mu\nu}$ to be the index-shifting metric, thereby taking care of occurrences of $\igc^{\mu\nu}$ implicitly through the use of covariant indices\footnote{It should be noted that this is merely a choice of notation, since using the classical metric as the true index-shifter will introduce errors, see section \ref{s:gauge}.}. Following this convention, the contributions to the curvature scalar can be compactly rewritten:
\bea
R^{(1)} &=& \left( \inabcl^\mu \inabcl^\nu - \igc^{\mu \nu} \dAcl - \irtcl^{\mu \nu}\right)\gamma_{\mu \nu} \label{eqn:fmsRicciS1} \ \text{, with} \ \\
\dAcl &=& \igc^{\rho \sigma} \nabcl_\rho \nabcl_\sigma \nn\\
R^{(2)}&=&\gamma_{\alpha \beta}\inabcl^\alpha \inabcl^\beta\gamma^{\rho}{}_{\rho}+\gamma_{\alpha \beta}\dAcl \gamma^{\alpha \beta}+\inabcl^{\alpha}\gamma_{\alpha \rho}\inabcl^\rho \gamma^{\mu}{}_{\mu} \nn\\
&&+\frac{3}{4}\inabcl^\alpha \gamma_{\rho}{}^{\mu}\nabcl_\alpha \gamma^{\rho}{}_{\mu} -\frac{1}{4}\inabcl^\rho \gamma^{\mu}{}_{\mu}\inabcl_\rho \gamma^{\alpha}{}_{\alpha}-\frac{1}{2}\inabcl^\alpha \gamma_{\rho}{}^{\mu}\inabcl^\rho \gamma_{\alpha \mu} \nn\\
&&-\inabcl^\alpha \gamma_{\alpha \rho}\inabcl^{\mu}\gamma^{\rho}{}_{\mu} -2\gamma_{\alpha \beta}\inabcl^\alpha \inabcl^\rho \gamma^{\beta}{}_{\rho} +\gamma_{\alpha \beta}\igc^{\alpha \lambda}\rcl_{\rho \lambda \epsilon}{}^{\beta}\gamma^{\epsilon \rho} \nn
\eea
The Riemann tensor $\rcl_{\rho \lambda \epsilon}{}^{\beta}$ that is occurring in the explicit version of $R^{(2)}$ is defined with the sign convention from \cite{wald2010general}, for the "classical" covariant  derivative \eqref{eqn:cov_deriv} and it encodes the curvature of the manifold related to the classical metric in the usual way\footnote{The term containing the Riemann tensor is part of the second order contribution since two of the covariant derivatives have been exchanged and it could be removed again, by reverting this, however, it will be useful to keep it this way.}.

With the expression for $R^{(1)}$ in \eqref{eqn:fmsRicciS1}, the lowest order contribution to the geon propagator can be expressed more explicitly. Assuming a convention where objects that are evaluated at event $\evB$ are denoted with primed indices (or merely a prime in the case of scalars),
\bea
D_R\left(\gdcl \left(\evA, \evB \right)\right) &=& \Big\langle \left( \inabcl^\mu \inabcl^\nu - \igc^{\mu \nu} \dAcl - R^{\mu \nu}_c\right)\gamma_{\mu \nu} \times \nn\\
&& \times \left( \inabcl^{\rho^\prime} \inabcl^{\sigma^\prime} - \igc^{\rho^\prime \sigma^\prime} \dAcl^\prime -R^{\rho^\prime \sigma^\prime}_c \right)\gamma_{\rho^\prime \sigma^\prime}\Big\rangle \nn\\
&=&\left( \inabcl^\mu \inabcl^\nu - \igc^{\mu \nu} \dAcl - R^{\mu \nu}_c\right) \times \nn\\
&& \times \left( \inabcl^{\rho^\prime} \inabcl^{\sigma^\prime} - \igc^{\rho^\prime \sigma^\prime} {\dAcl}^\prime -R^{\rho^\prime \sigma^\prime}_c \right)\langle \gamma_{\mu \nu} \gamma_{\rho^\prime \sigma^\prime}\rangle \ \text{.} \label{eqn:reducedGeonProp} \nn\\
\eea
This therefore expresses the geon propagator as a function of the propagator of the elementary graviton, consistent at the leading order in the FMS expansion\footnote{The previous result in \cite{Maas:2019eux} was of lower order as it did not include inverse fluctuations in the metric at the same order, but rather set them to zero.}.

Hence, it only remains to plug in the corresponding classical metric and the tree-level propagator of the elementary graviton in the Haywood gauge to obtain a result. The expressions for (anti-)de Sitter space-time are already too involved even at that level. Thus, we restrict here to $g^c$ being Minkowski metric only.

The required expressions for the case of a flat classical metric are derived in appendix \ref{a:prop}. In this case the Riemann tensor vanishes and the covariant derivatives become ordinary ones. The tree-level graviton propagator in a maximally symmetric space-time is determined by five tensor structures \cite{Allen:1986tt,Perez-Nadal:2009jcz},
\begin{align}
G_{\mu \nu;\lambda^\prime \epsilon^\prime}\left(\gdcl\right)&=g^c_{\mu\nu}g^c_{\lambda^\prime \epsilon^\prime}a(\gdcl)+\left(\prop_{\mu \lambda^\prime}\prop_{\nu \epsilon^\prime}+\prop_{\mu \epsilon^\prime}\prop_{\nu \lambda^\prime}\right)b(\gdcl) \nonumber\\& \ \ \ +\left(n_{\mu}n_{\lambda^\prime}\prop_{\nu \epsilon^\prime}+n_{\mu}n_{\epsilon^\prime}\prop_{\nu\lambda^\prime}+n_{\nu}n_{\lambda^\prime}\prop_{\mu \epsilon^\prime}+n_{\nu}n_{\epsilon^\prime}\prop_{\mu \lambda^\prime}\right)c(\gdcl) \nonumber\\
& \ \ \ +\left(n_{\mu}n_{\nu}g^c_{\lambda^\prime \epsilon^\prime}+g^c_{\mu\nu}n_{\lambda^\prime} n_{\epsilon^\prime}\right)d(\gdcl)+n_{\mu}n_{\nu}n_{\lambda^\prime} n_{\epsilon^\prime} e(\gdcl). \label{eqn:graviton_AJ_basis}
\end{align}
\no Herein $n_\mu=g_\mn^c n^\nu$ are determined by the tangent vectors of the (minimum length) geodesic connecting the two events and $\prop_{\mu \lambda^\prime}$ denotes the parallel propagator along this geodesic. The derivation in the Haywood gauge away from coincidence is reported for completeness in appendix \ref{a:prop}, and we recover the known results of \cite{Allen:1986tt,shao2003vacuum},
\begin{align}
-2 \kappa^2 G_{\mu \nu;\lambda^\prime \epsilon^\prime}\left(\gdcl\right)&=g^c_{\mu\nu}g^c_{\lambda^\prime \epsilon^\prime}\left(a_2+\frac{f_1}{16 \gdclB^2}-\frac{x_1}{2\gdclB^4}\right) \nonumber\\
& \ \ \ +\left(\prop_{\mu \lambda^\prime}\prop_{\nu \epsilon^\prime}+\prop_{\mu \epsilon^\prime}\prop_{\nu \lambda^\prime}\right)\left(b_2-\frac{13f_1}{16\gdclB^2}+\frac{x_1}{2\gdclB^4}\right) \nonumber\\& \ \ \ +\left(n_{\mu}n_{\lambda^\prime}\prop_{\nu \epsilon^\prime}+n_{\mu}n_{\epsilon^\prime}\prop_{\nu\lambda^\prime}+n_{\nu}n_{\lambda^\prime}\prop_{\mu \epsilon^\prime}+n_{\nu}n_{\epsilon^\prime}\prop_{\mu \lambda^\prime}\right)\frac{3f_1}{8 \gdclB^2} \nonumber\\
& \ \ \ +\left(n_{\mu}n_{\nu}g^c_{\lambda^\prime \epsilon^\prime}+g^c_{\mu\nu}n_{\lambda^\prime} n_{\epsilon^\prime}\right)\left(\frac{5f_1}{8\gdclB^2}+\frac{2x_1}{\gdclB^4}\right) \nonumber\\
& \ \ \ +n_{\mu}n_{\nu}n_{\lambda^\prime} n_{\epsilon^\prime} \left(-\frac{12x_1}{\gdclB^4}-\frac{f_1}{2\gdclB^2}\right).\nn
\end{align}
There remain two undetermined constant terms, which would be fixed at coincidence, but they do not play a role here, as they drop out for the geon propagator.

Applying the operator in \eqref{eqn:reducedGeonProp} to this expression for the graviton propagator leads to\footnote{Where $\Box_{ M} = \eta^\mn \partial_\mu \partial_\nu$ is the wave operator on the Minkowski space-time.
}
\begin{align}
D_{R\text{,tl}}\left(\gdcl \left( \evA, \evB \right)\right)&=-2i\kappa^2\Box^\prime_{ M}\Box_{ M}G^{\alpha}{}_{\alpha;}{}^{\lambda^\prime}{}_{\lambda^\prime} \nonumber\\
&= -i 6 \kappa^2\Box^\prime_{ M}\delta^{(4)} \left(x_{\evA} - x_{\evB}\right) \text{,}\nn
\end{align}
where the subscript \textbf{tl} indicates that we are only employing tree-level results for the fundamental graviton. The reduction of the graviton propagator, acted upon with a wave operator, is a consequence of it being a solution to the Green's function equation and can be verified with the results from appendix \ref{a:prop}. With this result we can finally provide an expression for the geon propagator in flat space to second order in the FMS expansion and tree-level in the Newton coupling,
\begin{align}
\big\langle R\left(\evA\right) R\left(\evB\right)\big\rangle_{\text{connected,tl}} = -i 6 \kappa^2\Box^\prime_{ M}\delta^{(4)} \left(x_{\evA} - x_{\evB}\right) + O\left(\gamma^3\right) \label{eqn:finalGeon}.
\end{align}
\no The result is thus a non-propagating local term\footnote{We note that the result \pref{eqn:finalGeon} is formally proportional to the 2-point vertex of a massless, scalar particle. While this is still not a propagating object, this could hint to a possible further subtlety. In the conventional quantum-field-theoretical setup of the FMS mechanism \cite{Maas:2017wzi}, the classical quantity is a space-time independent number, and thus allows to replace on both sides full Green's functions with connected and amputated ones. We have implicitly assumed this to be applicable in the present case. The result could also be interpreted as a hint that this is not possible, and that amputation needs to be done explicitly. This will require further study, and may be connected to questions concerning the interrelation between the LSZ formalism and the FMS mechanism \cite{Maas:2017wzi,MPS:unpublished}.}, as all others drop out. Thus, to this order in the FMS expansion around a classical Minkowski space-time there is no non-trivial propagating geon.

This would imply that for a flat classical metric $g^c$, which can only be a good approximation for a vanishing cosmological constant, the hope that the geon can act as dark matter \cite{Maas:2019eux} cannot be fulfilled. However, this does not exclude either a different result at not-vanishing cosmological constant, like is the case for our universe, nor non-scalar geonic dark matter. And, of course, there is also the possibility that our calculations are too rough. A possibility to improve upon the results would be to stay at leading order in the FMS expansion, but to employ a higher-order graviton propagator. Within the current scope, this requires asymptotic safety for a meaningful propagator, and such propagators are available in position space \cite{Bonanno:2021squ,Fehre:2021eob}. Another possibility would be to evaluate the geon propagator along the lines of \cite{Dai:2021fqb} in dynamical triangulation simulations, and compare it to the graviton propagator in the Haywood gauge in the same simulations.

\section{Three-point functions and static black-holes}\label{s:bh}

Interactions of quantum theories are encoded in vertices. While it is possible to apply the FMS expansion strictly also to vertices \cite{Maas:2018ska,Jenny:2022atm,Maas:2017wzi}, this would be an even more formidable endeavour than the geon propagator. A much more modest idea will therefore be pursued here, which nonetheless follows the same philosophy \cite{Maas:2019eux}.

Arguably one of the most iconic ideas about quantum gravity is Hawking evaporation of black holes. Likewise, the merging of particles into a black hole would be the corresponding reversed process. Given the scalar particle in our setup, we assume the existence of an operator $B$, which has overlap with a Schwarzschild black hole state \cite{Maas:2019eux}. Then, both processes are described by the same correlation function
\be
\la B(X)\phi(Y)\phi(Z)\ra\label{hr},
\ee
\no and differs only by the nature of the pairwise geodesic distances, if $X$ is in the future light-cone or the past light-cone of both $Y$ and $Z$ simultaneously. Other options would correspond to other physical processes. Of course, this information would require a separate calculation to actually answer.

Moreover, there is an important feature when it comes to black holes. Usually, black holes are considered to create a distinctive space-time metric. However, in a quantum gravity setup black holes themselves are observable objects. Hence, the influence of black holes on particles is described by correlation functions like \pref{hr}, rather than evaluating the trajectory of the scalar particles in a fixed space-time. This has an interesting consequence in our FMS setup: Because we fixed the space-time to be flat, which needs to be kept, the black hole exists on top of it. If we would be able to perform the full calculation, this would not have any impact, as the split is purely technical. In an actual approximate calculation, this is different.

We want to perform here a first exploratory investigation. We therefore will treat the black hole as an (almost) classical object. In that sense, we assume that the operator $B(X)$ becomes a classical field. This entails two consequences. The first is that the classical metric of a black hole does not treat all events equal. E.\ g., in Kruskal coordinates the curvature singularity is an exceptional structure. Moreover, the classical metric is a full space-time metric, and thus does not include creation or annihilation.

However, we want to be able to evaluate \pref{hr} also if the event $X$ is not coinciding with the curvature singularity. Thus, what happens is that by replacing the operator $B$ by a classical field $B_c$, we make it dependent not only on the event $X$, but also of special event structure $\La$ of the classical black hole, $B_c(\La,X)$.

In the spirit of our approach, when choosing a non-spinning black hole, the object $B_c$ needs to be diff\-eo\-mor\-phism-invariant and scalar. The scalar nature implies a Schwarzschild blackhole, and the simplest scalar, diff\-eo\-mor\-ph\-ism-invariant, non-vanishing objects associated with it are the Kretsch\-mann scalar $K$ and the second curvature invariant $I_2$. Without any further information, any linear combination of them is an equally good choice. Putting the curvature singularity $\La$  at the origin of coordinates on any spatial hypersurface of the metric $g^c$ then yields
\be
B_c(\La,X)=a\sqrt{K}+b\sqrt[3]{I_2} = \left(\sqrt{12}a-\sqrt[3]{12}b\right)\frac{r_S}{r_X}
\ee
\no where $r_S=2G_NM$ is the Schwarzschild radius, with $M$ the ``mass'' of the black hole. $r_X=r_X(\La,X)$ is a measure of the distance between the event $X$ and the curvature singularity, and will be specified below. Note that we made the assumption that the operator $B(x)$ behaves dimensionally as a curvature operator and that the prefactors are dimensionless mixing parameters, thus yielding the cube root of $I_2$. Since the functional dependence is the same in both terms, we arbitrarily set $a=1$ and $b=0$. After all, the prefactor would eventually be determined by the renormalization of the correlation function. Other choices of $B_c$ would not alter fundamentally the outcome, but would, of course, yield quantitatively quite different results.

While this issue is relatively straightforward, there is a second issue, which comes from the FMS setup. Because a fixed classical metric $g^c$ is chosen in the split \pref{fmssplit}, the usual idea that the black hole defines the complete metric outside its event horizon is not applicable. Since we consider the classical metric $g^c$ to be the (quantum) average behavior of the theory, the black hole has to be regarded as an excitation above the space-time described by $g^c$. If the theory would be linear, this could be obtained by adding the effects. However, as already our gauge condition is non-linear, this is not possible. We need therefore a different approach.

Since, after all, we are interested in evaluating the impact of the black hole on other particles in \pref{hr}, we will use the following approximation. Since $B_c$ is now a classical object, it can be moved outside the expectation value
\be
\la B(X)\phi(Y)\phi(Z)\ra\approx B_c(\La ,X)\la\phi(Y)\phi(Z)\ra.
\ee
\no Thus, the expectation value is merely the propagator of a scalar particle from event $Y$ to event $Z$. It is modified by a factor, which depends on the event $X$. Expanding furthermore the propagator to leading order in the FMS expansion, this yields
\be
B_c(\La,X)\la\phi(Y)\phi(Z)\ra = B_c (\La,\omega(\La,X)) D_{\phi}^c(\sigma^c(Y,Z))\label{fmsbhapprox},
\ee
\no with $\sigma^c$ being again the geodesic distance between $Y$ and $Z$ in the metric $g^c$, and $\omega$ is a, yet to be specified, information about the relation of the event $X$ and $\La$.

While it appears in \pref{fmsbhapprox} like $Y$ and $Z$ are now independent of $X$, this is not so. By making the black hole operator classical, the structure $\La$ introduced special events. The events $X$, $Y$ and $Z$ are now relative to this structure. Thereby the quantitative value of $D_\phi$ will become dependent on the proximity of $Y$ and $Z$ to $\La$. This is indeed relatively indirect.

This leaves the question of what the amplitude of $B_c$ is at this point, and thus of the definition of $\omega$. As $\omega$ should give an information about the relation of the event $X$ to $\La$ in such a way as to make this comparable to the geodesic distance $\sigma^c(Y,Z)$, we will define it in the following way. We take the geodesic distance in the classical metric $g^c$ on the same spatial hypersurface to the location of the Kruskal coordinate singularity given by $\La$, $s=\sigma^c(0,X)$. By our choice above, this is just the origin of the coordinate system on any spatial hypersurface. We then determine the event in Schwarzschild coordinates, which has the same geodesic distance to the black hole, $\sigma^{-1}_\text{BH}(s)$. Of course, by construction, this translation will fail, once $\sigma^{c}(0,X)$ reaches the event horizon radius, and therefore \pref{fmsbhapprox} will at best work outside the event horizon, and will probably not be a very good approximation very close to the event horizon. However, this also shows that the existence of the event horizon is not lost in this approximation.

Thus, our final expression is\footnote{Keep in mind that the geodesic distance entering the propagator is expanded around the classical metric as well.}
\be
\la B(X)\phi(Y)\phi(Z)\ra\approx B_c\left(\La,\sigma_{\text{BH}}^{-1}\left(\sigma^c(0,X)\right)\right)D_\phi^{c}(\sigma^{c}(Y,Z)).
\ee
\no It contains only known quantities: The black hole operator, the scalar propagator, and the geodesic distances in the black hole metric and in the classical metric around which we expand. It describes the interaction strength between a black hole and a scalar particle. By its very construction, it diverges towards the event horizon, but drops off once either the black hole is probed far away from its event horizon, or the propagation of the scalar particle is probed over very large distances. 

For example, when using flat Minkowski space-time for $g^c$, taking only the tree-level expression for $D_\phi$, and the Kretschmann scalar for the black hole operator $B_c$ with the black hole residing at the (spatial) origin, this yields in the Minkowski coordinate system
\begin{align} 
\la B(X)\phi(Y)\phi(Z)\ra &\approx \nonumber \\\sqrt{12}\frac{r_S}{r_X^3}\text{sign}((&x_Y-x_Z)^2) \frac{im^2}{4\pi^2}\frac{K_1\left(m\sqrt{(x_Y-x_Z)^2+i\epsilon}\right)}{m\sqrt{(x_Y-x_Z)^2+i\epsilon}}\label{coupling}\\
\sigma^c(0,X) = r_X\sqrt{1-\frac{r_S}{r_X}}&+\frac{1}{2}r_S\ln\left(2\frac{r_X}{r_S}\left(1+\sqrt{1-\frac{r_S}{r_X}}\right)-1\right)\label{strans}
\end{align}
\no where $x_Y$ and $x_Z$ are the geodesic distances to the origin. $\sigma^c(0, X)$ is again the desired geodesic distance from the origin on the spatial hypersurface in Minkowski coordinates to the point at which the black hole field should be evaluated (event $X$), and needs to be solved for $r_X$.  It thus needs to be translated by the implicit condition in \pref{strans}. Note that $\sigma^c(0, X)$ can become zero, which translates in Schwarzschild coordinates to a lower value of $r_X=r_S$, the Schwarzschild radius, as discussed previously. At large distances, where the Schwarzschild metric becomes asymptotically also Minkowski, $r_X\sim \sigma^c(0, X)$. Thus, the black hole operator is not evaluated at arbitrarily small distances in the Schwarzschild metric. Hence, \pref{coupling} is finite with respect to $r_X$, and can only diverge as a function of $(x_Y-x_Z)$. Additionally, it decays exponentially in any space-like direction, while only polynomial in time-like directions\footnote{Note that geometrically not all possible distance combinations of time-like and space-like are possible.}. 

Especially, if all distances are chosen equal and space-like, which corresponds to the symmetric configuration usually used in particle physics to define running couplings \cite{Bohm:2001yx}, we find that the coupling decays like $\exp(-mr)/r^{4}$. This corresponds to a screened interaction, similar to a screened Yukawa-type interaction, but decaying quicker than usual Newtonian interactions.

Summarizing, in this approximation we find that the interaction strength remains finite, and behaves in space-like and time-like directions as naively expected, especially at long distances. However, at very short distances, where the details of the black hole would necessarily be resolved, it is unlikely that our approximations would be reliable.

\section{Four-point functions and black-hole-particle \\scattering}\label{s:behss}

If we would attempt to discuss the scattering of a particle with a black hole, we would need to study a (connected) correlation function like
\be
\la B(X)B(Y)\phi(P)\phi(Q)\ra\nn,
\ee
\no and demand that the expectation values of the geodesic distances between $X$ and $Y$, and $P$ and $Q$ are future time-like oriented. The actual evolution of the scattering would proceed by requiring that the space-like distances between $X$ and $P$, and $Y$ and $Q$ shrink to zero, if the time-like distances go to zero, and grow again when the time-like distances grow again.

Using the same approach as in section \ref{s:bh} would imply that now two special worldlines appear, and the translation \pref{strans} would need to be done twice. Of course, this then starts to get problematic, as now the identification of either black-hole operator to be the initial or final state scattering partner becomes murky. Doing so simply by identifying the two black hole center worldlines will yield the expression 
\begin{align}
    \la B(X)B(Y)\phi(P)\phi(Q)\ra&\approx\\ 12\frac{r_S^2}{r_X^3r_Y^3}&\text{sign}((x_P-x_Q)^2)\frac{im^2}{4\pi^2}\frac{K_1\left(m\sqrt{(x_P-x_Q)^2+i\epsilon}\right)}{m\sqrt{(x_P-x_Q)^2+i\epsilon}}\nn.
\end{align}
\no Thus, except that the expression becomes more singular when both events $X$ and $Y$ are closer to the event horizon, little changes. Conversely, if $X$ and $Y$ are far away from the black hole center, little changes. Of course, this cannot be a very good approximation, as it depends on the special structure of the black holes. Thus, for this approximation to provide any realistic result, these events need to be close to the actual black holes. In this case the scattering cross section rises quickly, no matter from where the particle starts and ends. This is not surprising, as it describes a situation with a close approach to the black hole.

Thus, while the result is certainly plausible, it is not even qualitatively a suitable estimate. Still, it outlines how to approach particle-black-hole scattering in such a formalism. But it will require to get rid of the classical formulation of the black holes, and start to treat it as a genuine quantum operator. Though this immediately raises the question what a suitable operator would be for an even mesoscopic black hole. It appears therefore difficult for the moment to address such a process consistently.

\section{Summary}\label{s:sum}

We have explored herein how to extend FMS-augmented perturbation theory systematically to quantum gravity, based on the ideas put forward in \cite{Maas:2019eux}. One major step was the development of the necessary gauge conditions in a systematic way in section \ref{s:gauge}. As the present formalism is different from a background formulation, it is in a quantum gravity setting necessary to formulate the gauge fixing already without any reference to the classical part. This entails non-linearities, which are absent in a background-field formalism. We described one approach how to perturbatively deal with them. As this creates an infinite series of tree-level terms, this appears at first only worsen the problems in quantization. However, since in other FMS-augmented perturbative series the correct inclusion of gauge degrees of freedom diminished the problem \cite{Maas:2020kda}, it may actually be better than before. At the very least, this offers a new avenue to deal with the problem, which needs to be explored further.

This is, however, technically difficult as the examples showed. Especially, in flat space-time no non-trivial result is obtained. It may be very interesting to expand the calculation into non-flat space-time, to see whether differences arise. However, the exploration of 3-point functions illustrated that it very quickly becomes complicated to deal with problems beyond the Planck scale. This is partly due to the question of operators, but partly also because the question about relations of gravitationally-interacting particles with objects like a black hole is of similar complexity as of strongly-interacting particles with neutron stars. Here, further developments for the description of statistical ensembles of gravitons will be necessary, just like in the strong-interaction case.

Summarizing, we have developed a suitable framework to push FMS-augmented perturbation theory in canonical quantum gravity a step forward, and explored a number of sample applications. This should pave the way towards more developed calculations with less approximations, but it will be a very long way. To test, whether it would be worthwhile to walk it, it appears feasible \cite{Maas:2022bwc} to use dynamical triangulation simulations \cite{Ambjorn:2012jv,Loll:2019rdj} (or other approaches \cite{Schaden:2015wia,Catterall:2009nz,Asaduzzaman:2019mtx}), given the encouraging results in \cite{Dai:2021fqb} for the scalar particle. Given the great success in confirming the FMS mechanism in flat-space quantum field theory \cite{Maas:2017wzi,Maas:2020kda}, this appears a very promising avenue. Conversely, assuming the FMS approach to work, better accuracy could be obtained by staying at leading-order in the FMS approach, but use better results than tree-level for the gauge-dependent correlators, e.\ g.\ from asymptotic safety scenarios \cite{Bonanno:2021squ,Fehre:2021eob}.

\section*{Acknowledgments}

Research at the Perimeter Institute is supported in part by
Innovation, Science and Economic Development Canada,
and by the Province of Ontario, through the Ministry of
Colleges and Universities.

\appendix

\section{Geon propagator}\label{a:geon}

The terms that occur in the expressions for the FMS expansion of the curvature scalar $R$ are such that every contravariant index occurs on the inverse of the classical metric, i.e. $\igc^{\mu\nu}$, and therefore it is expedient to introduce a new convention. While for a proper mapping between co- an contravariant objects the full metric has to act as the index shifter and this is still the case in the FMS analysis, after the split the full metric does not occur anymore and as a convention it is useful to associate every contravariant index as with an implicit contraction of a covariant object with the classical metric that is $t^\mu = \igc^{\mu \nu} t_\nu$. This allows us to write the expressions after the split in a much more compact form, albeit much care is required when translating between objects that are expressed with the convention that $g^c_{\mu\nu}$ is the "index shifter" and expressions that contain $g_{\mu\nu}$ as the true index shifter. From now on we shall follow the new convention and whenever we revert to the original definition it will be explicitly pointed out.

Expressing first the Ricci scalar in terms of the metric only yields
\bea
R &=&- \tfrac{1}{4} g^{\alpha \beta } g^{\mu \nu } \Bigl(-4 \partial_{\nu }\partial_{\beta }g_{\alpha \mu } + 4 \partial_{\nu }\partial_{\mu }g_{\alpha \beta } + g^{\rho \sigma } \bigl(2 \partial_{\nu }g_{\beta \sigma } \partial_{\rho }g_{\alpha \mu } - 3 \partial_{\rho }g_{\alpha \mu } \partial_{\sigma }g_{\beta \nu } \nn\\
&&+ \partial_{\mu }g_{\alpha \beta } (\partial_{\nu }g_{\rho \sigma } - 4 \partial_{\sigma }g_{\nu \rho }) + 4 \partial_{\beta }g_{\alpha \mu } \partial_{\sigma }g_{\nu \rho }\bigr)\Bigr). \nn
\eea
Now we can identify the terms that are proportional to the Haywood gauge condition
\bea
R^H &=&- \tfrac{1}{4} g^{\alpha \beta } \biggl(4 H^{\mu } (\partial_{\beta }g_{\alpha \mu } -  \partial_{\mu }g_{\alpha \beta }) + g^{\mu \nu } \Bigl(-4 \partial_{\nu }\partial_{\beta }g_{\alpha \mu } + 4 \partial_{\nu }\partial_{\mu }g_{\alpha \beta } \nn\\
&&+ g^{\rho \sigma } \bigl(\partial_{\mu }g_{\alpha \beta } \partial_{\nu }g_{\rho \sigma } + \partial_{\rho }g_{\alpha \mu } (2 \partial_{\nu }g_{\beta \sigma } - 3 \partial_{\sigma }g_{\beta \nu })\bigr)\Bigr)\biggr). \nn
\eea
Setting $H^\mu$ to zero provides us with the Ricci scalar in Haywood gauge,
\bea
R^{H=0} &=& - \tfrac{1}{4} g^{\alpha \beta } g^{\mu \nu } \Bigl(-4 \partial_{\nu }\partial_{\beta }g_{\alpha \mu } + 4 \partial_{\nu }\partial_{\mu }g_{\alpha \beta } \nn\\
&&+ g^{\rho \sigma } \bigl(\partial_{\mu }g_{\alpha \beta } \partial_{\nu }g_{\rho \sigma } + \partial_{\rho }g_{\alpha \mu } (2 \partial_{\nu }g_{\beta \sigma } - 3 \partial_{\sigma }g_{\beta \nu })\bigr)\Bigr). \nn
\eea
Performing the split explicitly leads to the first order
\bea
R^{(1)}=&- \tfrac{1}{2} \igc^{\alpha \beta } \igc^{\mu \nu } \
\Bigl(-2 \partial_{\nu }\partial_{\beta }\gamma_{\alpha \mu } + 2 \
\partial_{\nu }\partial_{\mu }\gamma_{\alpha \beta } + \
\igc^{\rho \sigma } \bigl(\partial_{\mu }g^{c}{}_{\alpha
\beta } \partial_{\nu }\gamma_{\rho \sigma } \bigr.\Bigr.\nonumber\\
&\Bigl.\bigl.+ \partial_{\rho \
}g^{c}{}_{\alpha \mu } (2 \partial_{\nu }\gamma_{\beta \sigma } - 3 \
\partial_{\sigma }\gamma_{\beta \nu })\bigr)\Bigr), \nn
\eea
and second-order expressions
\bea
R^{(2)}&=&\tfrac{1}{4} \igc^{\alpha \beta } \igc^{\mu \nu } \igc^{\rho \sigma } \bigl(- \partial_{\mu }\gamma_{\alpha \beta } \partial_{\nu }\gamma_{\rho \sigma } + 4 \gamma_{\alpha \mu } \partial_{\nu }\partial_{\beta }g^{c}{}_{\rho \sigma } - 2 \partial_{\nu }\gamma_{\beta \sigma } \partial_{\rho }\gamma_{\alpha \mu } \nn\\
&&+ 3 \partial_{\rho }\gamma_{\alpha \mu } \partial_{\sigma }\gamma_{\beta \nu } - 8 \gamma_{\alpha \mu } \partial_{\sigma }\partial_{\nu }g^{c}{}_{\beta \rho } + 4 \gamma_{\alpha \mu } \partial_{\sigma }\partial_{\rho }g^{c}{}_{\beta \nu } \nn\\
&&+ \igc^{\tau \lambda } \gamma_{\alpha \mu } (-3 \partial_{\beta }g^{c}{}_{\rho \tau } \partial_{\nu }g^{c}{}_{\sigma \lambda } + \partial_{\beta }g^{c}{}_{\rho \sigma } \partial_{\nu }g^{c}{}_{\tau \lambda } + 2 \partial_{\rho }g^{c}{}_{\beta \nu } \partial_{\sigma }g^{c}{}_{\tau \lambda } 
\nn\\
&&- 6 \partial_{\lambda }g^{c}{}_{\nu \sigma } \partial_{\tau }g^{c}{}_{\beta \rho } + 4 \partial_{\nu }g^{c}{}_{\sigma \lambda } \partial_{\tau }g^{c}{}_{\beta \rho } + 2 \partial_{\sigma }g^{c}{}_{\nu \lambda } \partial_{\tau }g^{c}{}_{\beta \rho })\bigr). \nn
\eea
While this expression is useful in its own right, it is technically advantageous to recast them in a slightly different way \cite{Percacci:2017fkn}, which illuminates their structure better. This is achieved by collecting the terms, where only derivatives of the classical metric occur, in "classical" Christoffel symbols \eqref{eqn:classChrist} and defining the "classical" covariant derivative \eqref{eqn:classCD} based on these:
\bea
\chrisCl^{\alpha}{}_{\mu \nu} &=& \frac{1}{2} \igc^{\alpha \beta}\left(\partial_\mu g^c_{\beta \nu} + \partial_\nu g^c_{\mu \beta} - \partial_\beta g^c_{\mu \nu}\right), \label{eqn:classChrist}\\
\nabla^c_\mu v^\rho &=& \partial_\mu v^\rho + \chrisCl^{\rho}{}_{\mu \nu} v^\nu, \label{eqn:classCD}\\
\nabla^c_\rho g^c_{\mu \nu} &=& 0.
\eea
Furthermore, using the identity \eqref{eqn:gammaPartToCov} we can map specific combinations of partial derivatives of the $\gamma$-field to covariant derivatives of the $\gamma$-field. This combination of partial derivatives is the one occurring in the full Christoffel symbols of $g_{\mu\nu}$, after the split has been performed. Since the Ricci scalar is the trace part of the Ricci tensor, which is expressed in terms of Christoffel symbols, these are just the combinations that will occur in the end. Based on \eqref{eqn:gammaPartToCov} we define another auxiliary object \eqref{eqn:gammaChrist}, which corresponds to something similar to a "Christoffel" symbol for the $\gamma_{\mu\nu}$, with the partial derivatives replaced by \eqref{eqn:classCD}:
\bea
\partial_\mu \gamma_{\tau \nu} + \partial_\nu \gamma_{\mu \tau} -\partial_\tau \gamma_{\mu \nu} &=& \nabla^c_\mu \gamma_{\tau \nu} + \nabla^c_\nu \gamma_{\mu \tau} -\nabla^c_\tau \gamma_{\mu \nu} + 2\chrisCl^{\rho}{}_{\mu \nu}\gamma_{\rho \tau}, \label{eqn:gammaPartToCov} \\
\Theta^{\alpha}{}_{\mu \nu} &=& \frac{1}{2} \igc^{\alpha \tau}\left(\nabla^c_\mu \gamma_{\tau \nu} + \nabla^c_\nu \gamma_{\mu \tau} -\nabla^c_\tau \gamma_{\mu \nu}\right).\label{eqn:gammaChrist}
\eea
For the FMS expansion we consider the Ricci tensor at first and afterwards we take the trace to obtain the curvature scalar.
\bea
R_{\mu \nu} &=&\partial_\alpha \Gamma^{\alpha}{}_{\mu \nu}-\partial_\mu \Gamma^{\alpha}{}_{\alpha \nu}+\Gamma^{\beta}{}_{\mu \nu}\Gamma^{\alpha}{}_{\beta \alpha}-\Gamma^{\beta}{}_{\alpha \nu}\Gamma^{\alpha}{}_{\beta \mu} \nn\\
&=& R_{\mu \nu}^{(0)} + R_{\mu \nu}^{(1)} + R_{\mu \nu}^{(2)} + O\left(\gamma^3\right).
\eea
After the split the Christoffel symbols can be rearranged decomposed into the contributions at every order, using \eqref{eqn:classCD}, \eqref{eqn:classChrist}, \eqref{eqn:gammaPartToCov} and \eqref{eqn:gammaChrist}:
\bea
\Gamma^{\alpha}{}_{\mu \nu} = \chrisCl^{\alpha}{}_{\mu \nu} + \Theta^{\alpha}{}_{\mu \nu} - \igc^{\alpha \rho}\gamma_{\rho \sigma} \Theta^{\sigma}{}_{\mu \nu} + O\left(\gamma^3\right). \nn
\eea
This decomposition can then be inserted into the definition of the Ricci tensor and we immediately get the contribution to each order, with the zeroth order part trivially reducing to the classical Ricci tensor:
\bea
R_{\mu \nu}^{(0)} &=& R_{\mu \nu}^c, \nn\\
R_{\mu \nu}^{(1)} &=& \nabla^c_\alpha \Theta^{\alpha}_{\mu \nu} - \nabla^c_\mu \Theta^{\alpha}{}_{\alpha \nu}, \nn\\
R_{\mu \nu}^{(2)} &=& -\nabla^c_\alpha \left(\igc^{\alpha \tau}\gamma_{\tau \rho}\Theta^{\rho}{}_{\mu \nu}\right)+\nabla^c_\mu \left(\igc^{\alpha \tau}\gamma_{\tau \rho} \Theta^{\rho}{}_{\alpha \nu}\right)\nn\\
&&+\Theta^{\beta}{}_{\mu \nu}\Theta^{\alpha}{}_{\alpha \beta}-\Theta^{\beta}{}_{\alpha \nu}\Theta^{\alpha}{}_{\beta \mu}. \nn
\eea

\section{Scalar and graviton propagator}\label{a:prop}

In sections \pref{s:geon} and \pref{s:bh} it will be necessary to know the tree-level propagator of the graviton and the scalar particle, respectively. In addition, these will depend on the geodesic distance in any setup where events are treated equally. In this section, the necessary results will be collected.

Correlation functions involve the fields, which are functions of the events. Consequently, any correlation function can only depend on quantities, which respect all symmetries with respect to the events. This is an extension of usual translation invariance of non-gravitational quantum field theory. For propagators, the only such quantity available is the geodesic distance, which itself is again an expectation value \cite{Schaden:2015wia,Ambjorn:2012jv,Maas:2019eux}. At leading order \cite{Maas:2019eux}, this is just the classical geodesic distance. Thus, given the propagator $\la\phi(X)\phi(Y)\ra$, where capital letters will be used to label events, and $\sigma(X,Y)$ is the geodesic distance between both events, this yields
\be
\la\phi(X)\phi(Y)\ra=D_\phi(\la\sigma(X,Y)\ra)\nn.
\ee
\no Of course, for an actual calculation it is necessary to introduce coordinates, $x_X$. E.\ g.\ in flat space-time
\be
\la\sigma(X,Y)\ra=|x_X-x_Y|^2+\op(\gamma) = \gdcl(X,Y) +\op(\gamma), \nn
\ee
\no where the coordinates are determined on the fixed metric $g^c$. To this order, this is for $g^c(X)=\eta$, with $\eta$ the flat Minkowski metric, indeed the ordinary flat space distance.

Since $g^c$ is fixed, this would also allow to introduce a momentum space relative to $g^c$. This momentum space is then, of course, gauge-dependent. We will not do so here. However, we will introduce an effective mass as
\be
m(\sigma)=-\frac{1}{\sqrt{\sigma+i\epsilon}}\ln\left(\frac{2^\frac{7}{2}\pi^\frac{3}{2}\sigma}{(2\sigma+i\epsilon)^\frac{1}{4}}\Im D_\phi(\sigma)\right)\nn,
\ee
\no which approaches for large time-like $\sigma$ and $D_\phi$ the usual scalar flat-space propagator the flat-space mass.

For the scalar particle, the relevant term in the Lagrangian at two-point level reads
\be
\La=\frac{1}{2}\igc^{\mn}\pdm\phi\pdn \phi - \frac{1}{2}m^2\phi^2\nn.
\ee
\no Note that in this gauge all fluctuation terms from $\gamma^\mn$ are automatically interaction terms. Hence, the tree-level propagator is the one of a static space-time described by $g^c$. E.\ g., in flat space-time it is
\be
D_\phi(\gdcl)=\frac{\delta(\sigma^c)}{4\pi}+\frac{i}{4\pi^2}\frac{m}{\sqrt{2\gdcl+i\epsilon}}K_1\left(m\sqrt{2\gdcl+i\epsilon}\right),\label{pscalar}
\ee
\no where $K_1$ is the modified Bessel function of the first kind. Analogously, this yields the corresponding known tree-level propagator if $g^c$ is (anti-)de Sitter \cite{Parker:2009uva}.

The situation for the Graviton tree-level propagator is much more involved. After a lengthy calculation, a result for the quadratic\footnote{There is actually both a constant, i.\ e.\ $\gamma$-independent term and a term linear in $\gamma$. Neither can play a role in the following.} term in the Lagrangian arises, which is quite similar to the usual background method \cite{Percacci:2017fkn}, is obtained
\bea
\La&=&\frac{1}{4 \kappa }\gamma_{\alpha \beta}\left(\frac{1}{2}\igc^{\alpha \mu}\igc^{\beta \nu}\dAcl-\frac{1}{2}\igc^{\alpha \beta}\igc^{\mu \nu}\dAcl-\igc^{\beta \mu}\inabcl^\nu \inabcl^\alpha \right.\nn\\
&&\left.+2\igc^{\beta \mu}\rclB^{\alpha \nu}-\igc^{\mu\nu}\rclB^{\alpha \beta}-\frac{1}{2}\igc^{\beta \nu}\igc^{\mu \alpha}\rcl \right.
\nn\\
&&\left.+\frac{1}{4}\igc^{\mu \nu}\igc^{\alpha \beta}\rcl \right)\gamma_{\mu \nu} \nn\\
&=& \frac{1}{2 \kappa }\gamma_{\alpha \beta}\mathcal{D}^{\alpha \beta \mu \nu }\gamma_{\mu \nu},\nn
\eea
\no where quantities with a superscript 'c' contain only the classical metric $g^c$ and $\mathcal{D}$ has been symmetrized such that $\mathcal{D}^{\alpha \beta \mu \nu }=\mathcal{D}^{ \mu \nu \alpha \beta}$. This Lagrangian results in the following general Green's function (GF) equation,
\bea
\frac{1}{2 \kappa }\mathcal{D}_{\alpha \beta}{}^{\mu \nu } G_{\mu \nu;\lambda^\prime \epsilon^\prime} &=& \mathbb{I}_{\alpha \beta; \lambda^\prime \epsilon^\prime} \delta_4 \left(x_{\evA}, x_{\evB} \right) + \text{gauge terms, }
\eea
where the additional gauge terms need to be introduced to guarantee consistency with the chosen gauge condition, see below, and $\delta_4 \left(x_{\evA}, x_{\evB}\right)$ is the generalization of the Dirac Delta distribution to curved space-time, see \cite{Parker:2009uva}.

To make progress, it is now necessary to specify $g^c$. In case of a flat space-time, the resulting propagator, in de Donder (harmonic) gauge, is
\be
P_{\mu\nu\lambda \epsilon}(\gdcl)= \left(2\eta_{\mu \epsilon}\eta_{\nu \lambda} + 2\eta_{\mu \lambda}\eta_{\nu \epsilon} c-3\eta_{\mu \nu}\eta_{\lambda \epsilon}  \right)D_0(\gdcl),\nn
\ee
\no where $D_0$ is the zero-mass scalar propagator. We will work out the propagator in the Haywood gauge below.

The situation is substantially more complicated in de Sitter space-time. For maximally symmetric spaces one can find the decomposition shown in \eqref{eqn:graviton_AJ_basis}, which we will denote with
\begin{align}
    G_{\mu \nu;\lambda^\prime \epsilon^\prime}\left(\gdcl\right)&=\sum_{j=1}^5 O^{(j)}_{\mu \nu;\lambda^\prime \epsilon^\prime} f_j(\gdcl) \ \text{,} \nn
\end{align}
for brevity.\footnote{Note that in this decomposition the tensor on the r.h.s. of the propagator equation can be expressed as $\mathbb{I}_{\alpha \beta; \lambda^\prime \epsilon^\prime} = \frac{1}{2} O^{(2)}_{\mu \nu;\lambda^\prime \epsilon^\prime}$.} This decomposition makes use of the tangent vectors to the geodesics connecting the two events
in the correlation function as well as the parallel propagator which maps the tangent vector at one event to the tangent vector at
another event along the geodesic, $n^{\mu \prime} = \pi^{\mu \prime}{}_{\nu} n^\nu$. Rules for the algebraic manipulation of these objects have been derived in \cite{Allen:1986cmp,Allen:1986tt,Perez-Nadal:2009jcz}. Since
we are working in the lowest order FMS approximation, all of these quantities are defined w.r.t.\ the classical
metric $g^c$, which is why we can make use of maximal symmetry in simple choices of the latter.

Herein the functions $f_j(\gdcl)$ are solutions to the massive scalar equation on de Sitter space-time, and a last part, which is a solution to a more involved partial differential equation. The important statement is that the typical mass parameter is given to be of the order of the (tree-level) cosmological constant. This implies a very small mass at tree-level.

To demonstrate how this basis decomposition can be used to solve for the propagator in position space, we will apply this method to the graviton propagator in Haywood gauge for a classical Minkowski space-time.\footnote{Note that in Minkowski space-time the parallel propagator in the decomposition trivially reduces to the classical metric.}

So as to get a GF consistent with our gauge condition, we have to modify the r.h.s. of the GF equation with the two biscalar functions $\phi$ and $\xi$, which are defined as the solutions to the following differential equations:
\begin{align}
\Box_{\text{M}} \phi\left(\gdcl \right) &= \delta^{(4)},\quad\quad 
\Box_{\text{M}} \xi\left(\gdcl \right) = \phi\left(\gdcl \right)\nn
\end{align}
These can be reduced to ODE's in the geodesic distance, when applying the identity
\begin{align}
\Box_c f_j(\gdcl) = f_j^{\prime \prime}(\gdcl) + \frac{3}{\gdcl} f_j^\prime(\gdcl) \text{,}\nn
\end{align}
see \cite{Allen:1986tt}. In the following we will discuss the solution for the graviton propagator away from coincidence, which causes the Dirac Delta source to vanish and we only have to solve simple ODEs without any distributional source terms. Away from coincidence the equation for $\phi$ becomes:
\begin{align}
\Box_c \phi_j(\gdcl) = \phi_j^{\prime \prime}(\gdcl) + \frac{3}{\gdcl} \phi_j^\prime(\gdcl) = 0 \text{,}\nn
\end{align}
which has the general solution
\begin{align}
\phi\left(\gdcl\right)= \frac{f_1}{2 (\gdcl)^2} +f_2 \text{.}\nn
\end{align}
With this result the ODE for $\xi\left(\gdcl\right)$ becomes
\begin{align}
\xi_j^{\prime \prime}(\gdcl) + \frac{3}{\gdcl} \xi_j^\prime(\gdcl) = \phi = \frac{f_1}{(\gdcl)^2} +f_2 \text{,}\nn
\end{align}
which is solved by the function
\begin{align}
\xi\left(\gdcl\right)=\frac{-x_1}{2 (\gdcl)^2}+\frac{f_1}{4}\log (\gdcl) +\frac{f2}{8}(\gdcl)^2+x_2\text{.}\nn
\end{align}
The constants $f_1$ and $x_1$ are just numerical factors. The r.h.s. of the Green's function equation in terms of the basis bitensors $O^{(j)}$ with the scalar prefactors in terms of $\phi$ and $\xi$ reads,
\begin{align}
r.h.s. &=O^{(1)}\left(\psi+\chi\right) + O^{(2)} \left(\delta^{(4)} \left(x_{\evA} - x_{\evB}\right)+\psi-2\chi\right)\nonumber \\
& \ \ \ +O^{(3)} \left(-\sigma+\omega\right)+O^{(4)}_{\text{L}}\left(\sigma+\omega\right)+O^{(4)}_{\text{R}}\sigma +O^{(5)}\tau  \text{,} \label{eqn:rhsAJbasis}
\end{align}
where we have suppressed the indices $\lbrace \alpha \beta; \lambda^\prime \epsilon^\prime \rbrace$ on each basis bitensor\footnote{Here we had to introduce the auxiliary tensor structure $O^{(4)}_{\text{L}}=n_\alpha n_\beta g^c_{\lambda^\prime \epsilon^\prime}$ and $O^{(4)}_{\text{R}}=g^c_{\alpha \beta}n_{\lambda^\prime}n_{\epsilon^\prime}$.}. And we have introduced the following short-hand notation
\begin{align}
\omega &\equiv \phi^{\prime \prime}-(\gdcl)^{-1}\phi^\prime \text{,} &\psi \equiv (\gdcl)^{-2}\xi^{\prime\prime}-(\gdcl)^{-3}\xi^\prime\text{,}\nonumber\\
\chi &\equiv (\gdcl)^{-1}\phi^\prime \text{,} &\sigma \equiv (\gdcl)^{-1} \xi^{\prime\prime\prime}-3(\gdcl)^{-2}\xi^{\prime\prime}+3(\gdcl)^{-3}\xi^\prime \text{,}\nonumber\\
\tau &\equiv \xi^{\prime\prime\prime\prime}-(\gdcl)^{-1}\xi^{\prime\prime\prime}-5\sigma \text{.}\nn
\end{align}
The l.h.s.\ of the GF equation derives from the following two expressions:
\begin{align}
\frac{1}{2}\left(O^{(2)}_{\alpha \beta}{}^{\mu \nu} -\eta_{\alpha \beta}\eta^{\mu \nu}\right)G_{\mu \nu;\lambda^\prime \epsilon^\prime} &\refeq{eqn:graviton_AJ_basis}\frac{1}{2}\left(\left(-3 a-2b-d\right)O^{(1)}+bO^{(2)}+cO^{(3)}\right.\nn\\
&\left. \ \ \ \ \ \ +d O^{(4)}_{\text{L}} +\left(4c-3d-e\right)O^{(4)}_{\text{R}}+eO^{(5)}\right) \text{,}\nn
\end{align} 
and using the manipulation rules from \cite{Allen:1986cmp,Allen:1986tt},
\begin{align}
\Box_{\text{M}} \frac{1}{2}&\left(O^{(2)}_{\alpha \beta}{}^{\mu \nu}-\eta_{\alpha \beta}\eta^{\mu \nu}\right)G_{\mu \nu;\lambda^\prime \epsilon^\prime} =\nn\\ =&\frac{1}{2}\left(O^{(1)} \left(-3 \Box_{\text{M}}a-2\Box_{\text{M}}b -\left(\Box_{\text{M}}+4(\gdcl)^{-2}\right)d +8(\gdcl)^{-2}c-2(\gdcl)^{-2}e\right)\right. \nn\\
&\left. \ \ \ +O^{(2)} \left(\Box_{\text{M}}b-4(\gdcl)^{-2}c\right) \right.\nn\\
&\left. \ \ \ +O^{(3)} \left(\left(\Box_{\text{M}}-8(\gdcl)^{-2}\right)c-2(\gdcl)^{-2}e\right) \right.\nonumber\\
&\left. \ \ \ +O^{(4)}_{\text{L}} \left(\left(\Box_{\text{M}}-8(\gdcl)^{-2}\right)d+2(\gdcl)^{-2}e \right) \right.\nonumber\\
&\left. \ \ \ +O^{(4)}_{\text{R}} \left(4\left(\Box_{\text{M}}-8(\gdcl)^{-2} \right)c-3\left(\Box_{\text{M}}-8(\gdcl)^{-2}\right)d \right.\right.\nonumber\\
&\left.\left. \ \ \ -\left(\Box_{\text{M}}-10(\gdcl)^{-2}\right)e\right)+O^{(5)}\left(\Box_{\text{M}}-24(\gdcl)^{-2}\right)e\right)\text{.} \label{eqn:minGFE}
\end{align}
where the indices of the basis objects $O^{(j)}$ are suppressed on the r.h.s., since they are always of the form $\lbrace \alpha \beta; \lambda^\prime \epsilon^\prime \rbrace$. Furthermore, also the argument $\gdcl$ of the scalar coefficient functions is suppressed for brevity. 

Now we can combine \eqref{eqn:minGFE} and \eqref{eqn:rhsAJbasis} in the Green's function equation and we can treat the prefactor for each basis object $O^{(j)}$ as a separate scalar differential equation,
\begin{align}
O^{(1)}&: 3 \Box_{\text{M}}\tilde{a}+2\Box_{\text{M}}\tilde{b} + \left(\Box_{\text{M}}+4(\gdcl)^{-2}\right)\tilde{d} -8(\gdcl)^{-2}\tilde{c}+2(\gdcl)^{-2}\tilde{e} = \psi+\chi, \nonumber\\
O^{(2)}&: -\Box_{\text{M}}\tilde{b}+4(\gdcl)^{-2}\tilde{c} = \delta^{(4)} \left(x_{\mathscr{P}} - x_{\mathscr{Q}}\right)+\psi-2\chi, \nonumber\\
O^{(3)}&: -\left(\Box_{\text{M}}-8(\gdcl)^{-2}\right)\tilde{c}+2(\gdcl)^{-2}\tilde{e} = -\sigma+\omega, \nonumber\\
O^{(4)}_{\text{L}}&: -\left(\Box_{\text{M}}-8(\gdcl)^{-2}\right)\tilde{d}-2(\gdcl)^{-2}\tilde{e} = \sigma+\omega, \nonumber\\
O^{(4)}_{\text{R}}&: -4\left(\Box_{\text{M}}-8(\gdcl)^{-2} \right)\tilde{c}+3\left(\Box_{\text{M}}-8(\gdcl)^{-2}\right)\tilde{d}+\left(\Box_{\text{M}}-10(\gdcl)^{-2}\right)\tilde{e} = \sigma, \nonumber\\
O^{(5)}&: -\left(\Box_{\text{M}}-24(\gdcl)^{-2}\right)\tilde{e} = \tau \text{,}\nn
\end{align}
where the scalar functions with a tilde are defined by the original functions through the following rescaling
\begin{align}
\tilde{f}_j=-\frac{1}{2\kappa^2}f_j \text{.}\nn
\end{align}
This system of ordinary differential equations is in fact soluble and the solutions to the five scalar coefficient functions $\lbrace a(\gdcl),b(\gdcl),c(\gdcl),d(\gdcl),e(\gdcl)\rbrace$ away from coincidence are worked out in the following. First, we can rewrite the equations to
\begin{align}
\Box_M\tilde{a}&=\frac{1}{3}\left(-2\Box_M\tilde{b}-\Box_M\tilde{d}+8(\gdcl)^{-2}\tilde{c}-4(\gdcl)^{-2}\tilde{d}-2(\gdcl)^{-2}\tilde{e}\right.\nn\\
&\left.\phantom{8(\gdcl)^{-2}} +\psi+\chi\right),\nn\\
\Box_M\tilde{b}&=4(\gdcl)^{-2}\tilde{c}-\psi+2\chi,\nn\\
\Box_M\tilde{c}-8(\gdcl)^{-2}\tilde{c}&=2(\gdcl)^{-2}\tilde{e}+\sigma-\omega,\nn\\
\Box_M\tilde{d}-8(\gdcl)^{-2}\tilde{d}&=-2(\gdcl)^{-2}\tilde{e}-\sigma-\omega,\nn\\
\Box_M\tilde{e}-24(\gdcl)^{-2}\tilde{e}&=-\tau \text{.}\nn\\
\label{eqn:dal_b}
\end{align}
If we solve the equations from bottom to top, starting with $\tilde{e}$, the source terms on the r.h.s. are always known. Notice that the Dirac delta in equation 2 of the system in \eqref{eqn:dal_b} has been omitted in this step. The sixth equation related to the tensor structure of $O^{(4)}_{\text{L}}$ can be shown to be satisfied automatically by the obtained solution,
\begin{align}
\tilde{a}\left(\gdcl\right) &= a_2+\frac{e_1}{48(\gdcl)^6}-\frac{d_1}{2(\gdcl)^4}+\frac{x_1}{2(\gdcl)^4}-\frac{a_1}{2(\gdcl)^2}-\frac{d_2}{2}(\gdcl)^2+\frac{e_2}{48}(\gdcl)^4,\nn\\
\tilde{b}\left(\gdcl\right) &= b_2+\frac{e_1}{48(\gdcl)^6}+\frac{c_1}{2(\gdcl)^4}+\frac{x_1}{2(\gdcl)^4}-\frac{b_1}{2(\gdcl)^2}+\frac{c_2}{2}(\gdcl)^2+\frac{e_2}{48}(\gdcl)^4,\nn\\
\tilde{c}\left(\gdcl\right) &= c_2 (\gdcl)^2+\frac{c_1}{(\gdcl)^4}+\frac{e_1+3f_1(\gdcl)^4+e_2(\gdcl)^{10}}{8(\gdcl)^6},\nn\\
\tilde{d}\left(\gdcl\right) &= d_2 (\gdcl)+\frac{d_1}{(\gdcl)^4}+\frac{-e_1+5f_1(\gdcl)^4-e_2(\gdcl)^{10}}{8(\gdcl)^6},\nn\\
\tilde{e}\left(\gdcl\right) &= e_2(\gdcl)^4+\frac{e_1}{(\gdcl)^6}-\frac{24x_1+f_1(\gdcl)^2}{2(\gdcl)^4} \ \text{,}\nn \\
\label{eqn:SolScalCoeff}
\end{align}
which is also expected, since there are only 5 independent tensor structures and the 6th
only appeared, because we split the tensor $O^{(4)}$ into two parts during the calculation.
For the frame-adapted approach we also required that the transversality condition is fulfilled. Acting the derivative on any index of $G_{\mu \nu;\lambda^\prime \epsilon^\prime}$ and using the ansatz \eqref{eqn:graviton_AJ_basis} this provides three conditions \cite{Allen:1986tt},
\begin{align}
\tilde{a}^\prime+\tilde{d}^\prime+3(\gdcl)^{-1}\tilde{d}-2(\gdcl)^{-1}\tilde{c} &= 0 \text{,}\nn\\
\tilde{e}^\prime+3(\gdcl)^{-1}\tilde{e}+\tilde{d}^\prime-2(\gdcl)^{-1}\tilde{d}-2\tilde{c}^\prime+4(\gdcl)^{-1}\tilde{c} &= 0 \ \text{and}\nn\\
-\tilde{b}^\prime-(\gdcl)^{-1}\tilde{d}-\tilde{c}^\prime+4(\gdcl)^{-1}\tilde{c} &=0 \text{.}\nn
\end{align}
Upon insertion of the solution \eqref{eqn:SolScalCoeff} for the scalar coefficients, these provide a set of eight constraints for the yet undetermined coefficients,
\begin{align}
-2c_1+d_1-2x_1 &=0, \nn\\
a_1+\frac{f_1}{8} &=0, \nn\\
-2c_2+4d_2 &=0, \nn\\
e_2 &=0, \nn\\
e_1 &=0, \nn\\
-24b_1-39f_1 &=0, \nn\\
c_2-d_2 &=0, \nn\\
240c_1-24d_1+48x_1 &=0. \nn
\end{align}
Thus the 12 coefficients for the general solution can be reduced to 4 remaining coefficients,
\begin{align}
\tilde{a}\left(\gdcl\right) &= a_2+\frac{f_1}{16 (\gdcl)^2}-\frac{x_1}{2(\gdcl)^4},\nn\\
\tilde{b}\left(\gdcl\right) &= b_2-\frac{13f_1}{16(\gdcl)^2}+\frac{x_1}{2(\gdcl)^4},\nn\\
\tilde{c}\left(\gdcl\right) &= \frac{3f_1}{8 (\gdcl)^2},\nn\\
\tilde{d}\left(\gdcl\right) &= \frac{5f_1}{8(\gdcl)^2}+\frac{2x_1}{(\gdcl)^4},\nn\\
\tilde{e}\left(\gdcl\right) &= -\frac{12x_1}{(\gdcl)^4}-\frac{f_1}{2(\gdcl)^2}.\nn
\end{align}
With this the graviton propagator away from coincidence becomes:
\begin{align}
-2 \kappa^2 G_{\mu \nu;\lambda^\prime \epsilon^\prime}\left(\gdcl\right)&=g^c_{\mu\nu}g^c_{\lambda^\prime \epsilon^\prime}\left(a_2+\frac{f_1}{16 \gdclB^2}-\frac{x_1}{2\gdclB^4}\right) \nonumber\\
& \ \ \ +\left(\prop_{\mu \lambda^\prime}\prop_{\nu \epsilon^\prime}+\prop_{\mu \epsilon^\prime}\prop_{\nu \lambda^\prime}\right)\left(b_2-\frac{13f_1}{16\gdclB^2}+\frac{x_1}{2\gdclB^4}\right) \nonumber\\& \ \ \ +\left(n_{\mu}n_{\lambda^\prime}\prop_{\nu \epsilon^\prime}+n_{\mu}n_{\epsilon^\prime}\prop_{\nu\lambda^\prime}+n_{\nu}n_{\lambda^\prime}\prop_{\mu \epsilon^\prime}+n_{\nu}n_{\epsilon^\prime}\prop_{\mu \lambda^\prime}\right)\frac{3f_1}{8 \gdclB^2} \nonumber\\
& \ \ \ +\left(n_{\mu}n_{\nu}g^c_{\lambda^\prime \epsilon^\prime}+g^c_{\mu\nu}n_{\lambda^\prime} n_{\epsilon^\prime}\right)\left(\frac{5f_1}{8\gdclB^2}+\frac{2x_1}{\gdclB^4}\right) \nonumber\\
& \ \ \ +n_{\mu}n_{\nu}n_{\lambda^\prime} n_{\epsilon^\prime} \left(-\frac{12x_1}{\gdclB^4}-\frac{f_1}{2\gdclB^2}\right).\nn
\end{align}
There remain four undetermined coefficients in the solution. Since we employed as an approximation that the fluctuation field should fall of fast enough for large distances, the propagator should also vanish for large enough distances. However, the two constant terms contradict that (at least in Minkowski space) and should thus also vanish. As for the constants $f_1$ and $x_1$, these are determined through the full Greens function equation, which also considers the coincidence term. Consequently, the current solution is still missing the contributions from the  the ultra-local terms.

\bibliographystyle{bibstyle}
\bibliography{bib}


\end{document}